\documentclass[%
 reprint,
 aps,
]{revtex4-1}

\usepackage{amsmath,amssymb}
\usepackage{graphicx}
\usepackage{dcolumn}
\usepackage{bm}
\usepackage{xcolor}
\usepackage{siunitx}
\usepackage{subcaption}
\usepackage{hyperref}

\begin{document}

\preprint{APS/123-QED}

\title{Electrically Controlled Two-Dimensional Electron-Hole Fluids}
\author{Yongxin Zeng and A.H. MacDonald} 
\affiliation{Department of Physics, University of Texas at Austin, Austin, TX 78712}
\date{\today}

\begin{abstract}
We study the electronic properties of dual-gated electron-hole bilayers in which the 
two layers are separated by a perfectly opaque tunnel barrier. 
Combining an electrostatic and thermodynamic analysis with mean-field theory estimates of 
interacting system chemical potentials, we explain
the dependence of the electron and hole densities on the two gate voltages. 
Because chemical potential jumps occur for both electrons and holes at neutrality, 
there is a finite area in gate voltage parameter space over which electron and hole densities are equal.  
In that regime the electron-hole pair density depends only on the 
sum of the two gate voltages.  We are able to explain a recent experimental study of 
electrically controlled bilayers by allowing for interlayer tunneling and using a non-equilibrium 
steady-state transport picture. 
\end{abstract}

\maketitle


\section{Introduction}
\begin{figure}
\centering
\includegraphics[width=\linewidth]{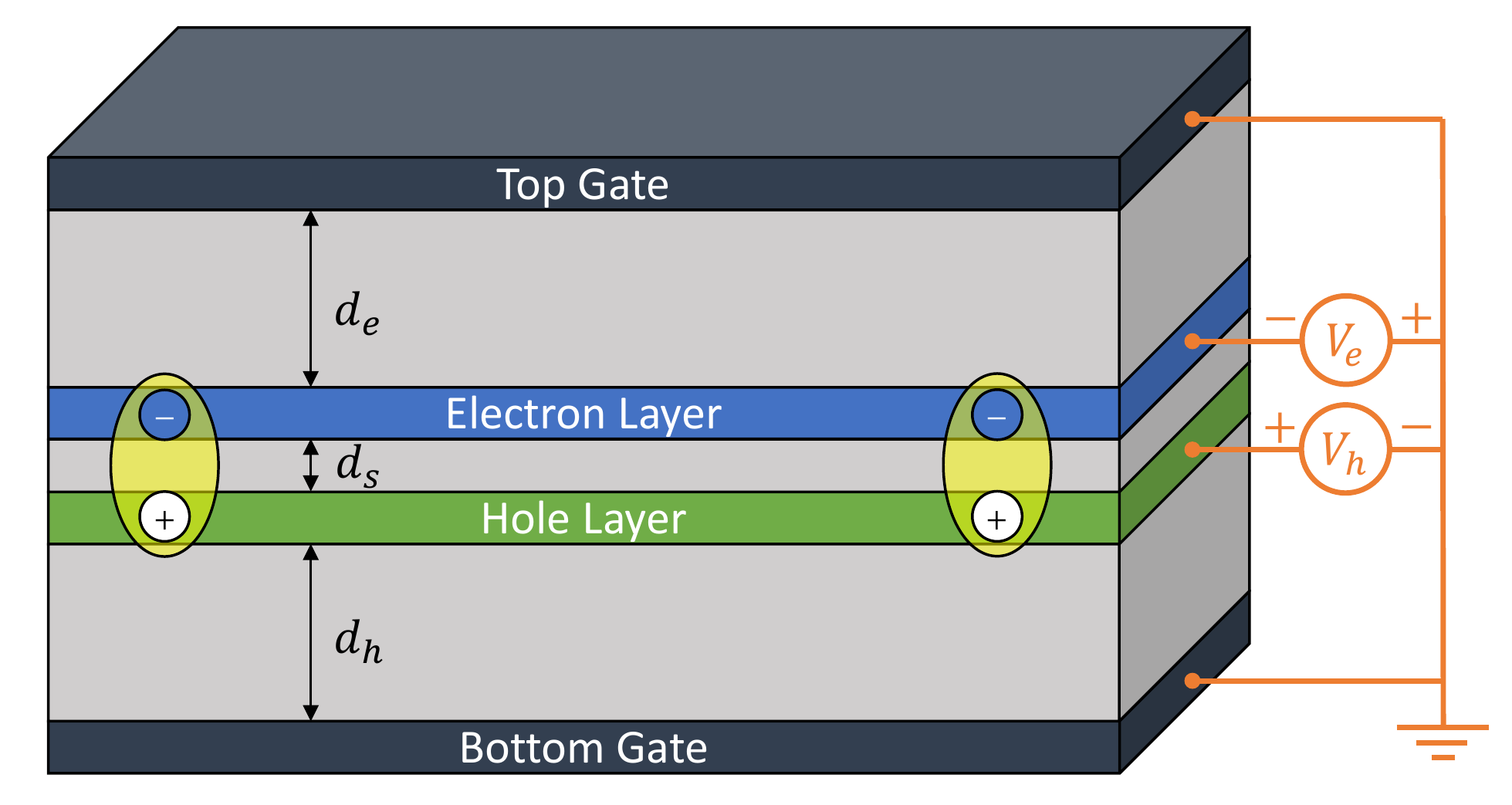}
\caption{Electrically controlled electron-hole fluids. We consider a dual gated geometry with both gates held at the same voltage.  
The electron and hole layers are contacted
by electrodes with voltages $-V_e$ and $+V_h$ respectively relative to the gate.}
\label{setup}
\end{figure}

Recent progress \cite{wang2019evidence,rivera2015observation,rivera2016valley,mak2016photonics,kim2017observation,burg2018strongly,wang2018electrical,jauregui2019electrical,ciarrocchi2019polarization,tran2019evidence,jin2019observation,seyler2019signatures,regan2020mott,tang2020simulation} in fabricating and processing two-dimensional (2D) semiconductor 
multilayers has opened up new opportunities to realize and study the properties of 2D electron-hole systems. 
Provided that the tunneling between layers is negligible, a bias voltage applied to a bilayer
with a spatially indirect band gap, 
like the one in Fig.~\ref{setup},
simply adjusts the effective band gap.   
Above a threshold bias voltage the 
effective band gap is reduced to zero so that 
electrons are induced in the conduction band of 
one 2D semiconductor layer, and holes in the valence band of the adjacent layer.
In the limit of low electron and hole densities, a regime of bias and 
gate voltages exists in which the electrons and holes pair into excitonic bound states, and separate electrical 
contacts can then be used to adjust the exciton chemical 
potential \cite{xie2018electrical}.  In the general case,
however, the electron and hole densities are unequal, 
allowing unprecedented access to systems of 
fermions with attractive mutual interactions, whose densities are unbalanced \cite{tiene2020extremely,subasi2010stability,parish2011supersolidity,varley2016structure}.
In this Article we outline a rigorous framework that relates the dependence of electron and 
hole densities in the semiconductor bilayers on bias and gate voltages to thermodynamic 
properties of the interacting 
electron-hole fluid, focusing first on the case of perfectly opaque barriers 
between the active layers and then discussing transport and electronic properties in the case 
where charge leakage between layers plays 
an important role.  

Our paper is organized as follows.  
Section \ref{sec:electrostatics} presents a formal analysis of the perfect barrier limit in terms of relevant 
thermodynamic properties of the interacting electron-hole fluid.  In Section \ref{sec:mft} we use a mean-field theory 
of the electron-hole fluid to obtain some concrete but approximate results which illustrate important 
key qualitative features, in particular the consequences of the singularities that electron-hole pairing 
produces in thermodynamic properties. In Section \ref{sec:control} we combine the results from the previous two sections and show how the carrier densities in the two layers are controlled by the gate voltages. 
We find that there is a finite area region in the gate-control parameter space 
where the electron and hole densities are exactly equal, which is
especially interesting for the study of exciton physics. Section \ref{sec:leak} generalizes the 
analysis to the case of non-zero leakage currents and provides an explanation for the 
transport characteristics observed in a recent experiment \cite{wang2019evidence}.  
Finally in Section \ref{sec:discussion} we conclude with a discussion of the prospects for 
realizing new physics in these systems.

\section{Dual Gated Electron-Hole Fluids} \label{sec:electrostatics}

We assume a dual gated structure like that illustrated in Fig.~\ref{setup}.  For definiteness we 
will focus on the case in which the two gates are physically identical and held at the same voltage.
We choose the Fermi energy of the gates as the zero of energy, and also set the zero of the electric potential 
at the gate planes.  We assume that the voltages on the contacts to the electron and hole layers, $-V_e$ and $+V_h$, can 
be varied at will, and that no current can flow between layers.  Under that circumstance the contacts 
on the two layers act like reservoirs whose voltages fix the electrochemial potentials of
the electrons in those layers.  We also assume that $V_e >0$ and $V_h >0$ so that the large voltages 
populate the conduction band in the electron layer and the valence band in the hole layer, allowing us to 
disregard the remote band degrees of freedom in both layers.

Starting from neutrality and increasing $V_e$, the electron layer begins to charge
when $V_e$ reaches the threshold voltage at which its electrochemical potential reaches the bottom of the conduction band $E_c$, whereas the hole layer begins to charge when its electrochemical potential reaches 
the top of the valence band, $E_v$.  We focus on the regime in which both layers are charged, 
which we refer to as the {\it p-n} regime. 
Applying the Poisson equation 
we obtain the following relationship between the electric potentials $-\phi_e$
and $+\phi_h$ and the free carrier densities $n$ and $p$ in the electron and hole layers: 
\begin{equation}
\begin{split} \label{eq:np_phi}
\frac{4\pi ne^2}{\epsilon} &= \frac{\phi_e}{d_e} + \frac{\phi_e+\phi_h}{d_s} \\
\frac{4\pi pe^2}{\epsilon} &= \frac{\phi_h}{d_h} + \frac{\phi_e+\phi_h}{d_s},
\end{split}
\end{equation}
where $d_e$ and $d_h$ are the distances from top and bottom gates to the electron and hole layers, and $d_s$ 
is the spacing between electron and hole layers.
Note that we have absorbed a factor of electron charge $e$ 
in the definition of $\phi_{e/h}$ and $V_{e/h}$ so that they 
have units of energy.  The inverse of this linear relationship is 
\begin{equation}
\begin{split} \label{eq:phi_np}
\phi_e(n,p) &= \frac{4\pi e^2 d_e}{\epsilon} \;\frac{(d_h+d_s)n-d_h p}{d_e+d_h+d_s} \\
\phi_h(n,p) &= \frac{4\pi e^2 d_h}{\epsilon} \;\frac{(d_e+d_s)p-d_e n}{d_e+d_h+d_s}
\end{split}
\end{equation}
Assuming that $d_e, d_h \gg d_s$, as is typical in experiment \cite{wang2019evidence},
we see that $n$ and $p$ will typically be nearly equal in magnitude, {\it i.e.} that this gating geometry 
favors nearly balanced electron-hole fluids but allows for large values of the 
individual layer densities:
\begin{equation}
n \approx p \approx \frac{\epsilon}{4\pi e^2}\frac{\phi_e+\phi_h}{d_s}
\end{equation}
Due to the small interlayer distance, the electric field between the two layers is typically
much larger than the electric field between the active layers and the gates.

The dependence of the carrier densities on the bias voltages applied to the active 
layers is determined by the following pair of non-linear equations:
\begin{equation}
\begin{split} \label{eq:equil}
\phi_e(n,p) + \mu_e(n,p) &= V_e-E_c \\
\phi_h(n,p) + \mu_h(n,p) &= V_h-E_v
\end{split}
\end{equation}
Note that $E_c$ and $E_v$ in these equations refer to the conduction and valence band edges
in the absence of carriers and are defined relative to the Fermi level of the gates.
The left-hand sides of these equations are functions of $n$ and $p$ with 
$\phi_{e}$ and $\phi_{h}$ given by the electrostatic equations \eqref{eq:phi_np}, and 
$\mu_e$ and $\mu_h$ determined by the many-body physics of the electron-hole fluid.
In Eq.~\eqref{eq:equil} $\mu_{e}$ and $\mu_{h}$ are defined as the electrochemical 
potentials of the electron and hole layers relative to their local band extrema.  
Experiments which measure 
$n$ and $p$ as a function of $V_e$ and $V_h$ therefore provide a valuable thermodynamic
probe of electron-hole fluid properties.  In the following section we 
discuss mean-field theory estimates for $\mu_e(n,p)$ and $\mu_h(n,p)$
that provide valuable insight into expected properties.
In general Eqs.~\ref{eq:equil} define two lines in $(n,p)$ space that intersect at the
equilibrium carrier densities.  As we discuss below, however, singularities along the 
$n=p$ line associated with the formation of excitonic bound states require special considerations.

\section{Mean-Field Theory of the Unbalanced Electron-Hole Fluid} \label{sec:mft}

\begin{figure*}
\centering
\begin{subfigure}{0.32\linewidth}
\includegraphics[width=\linewidth]{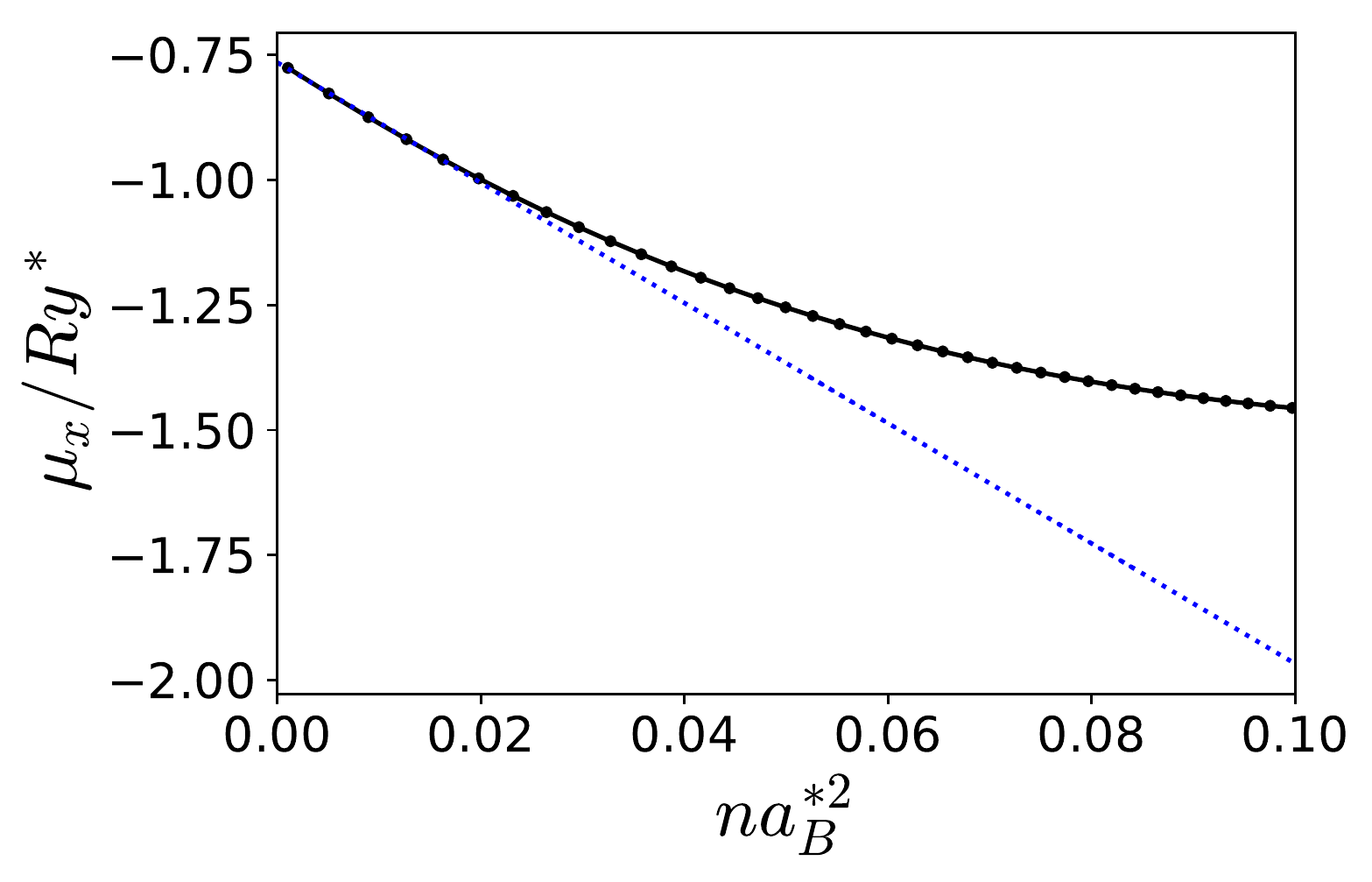}
\subcaption{} \label{fig:mux_n=p}
\end{subfigure}
\begin{subfigure}{0.32\linewidth}
\includegraphics[width=\linewidth]{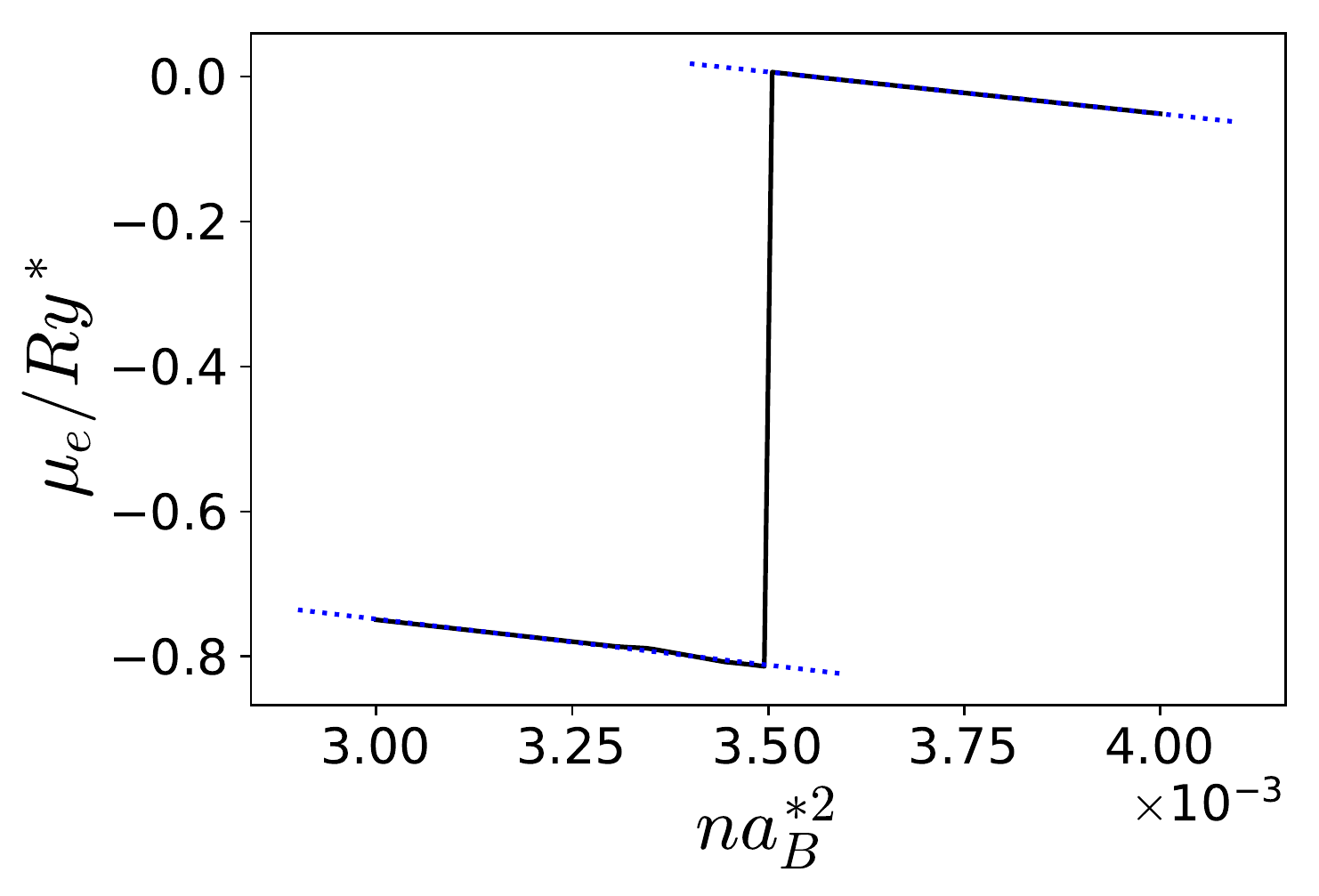}
\subcaption{} \label{fig:mue_n}
\end{subfigure}
\begin{subfigure}{0.32\linewidth}
\includegraphics[width=\linewidth]{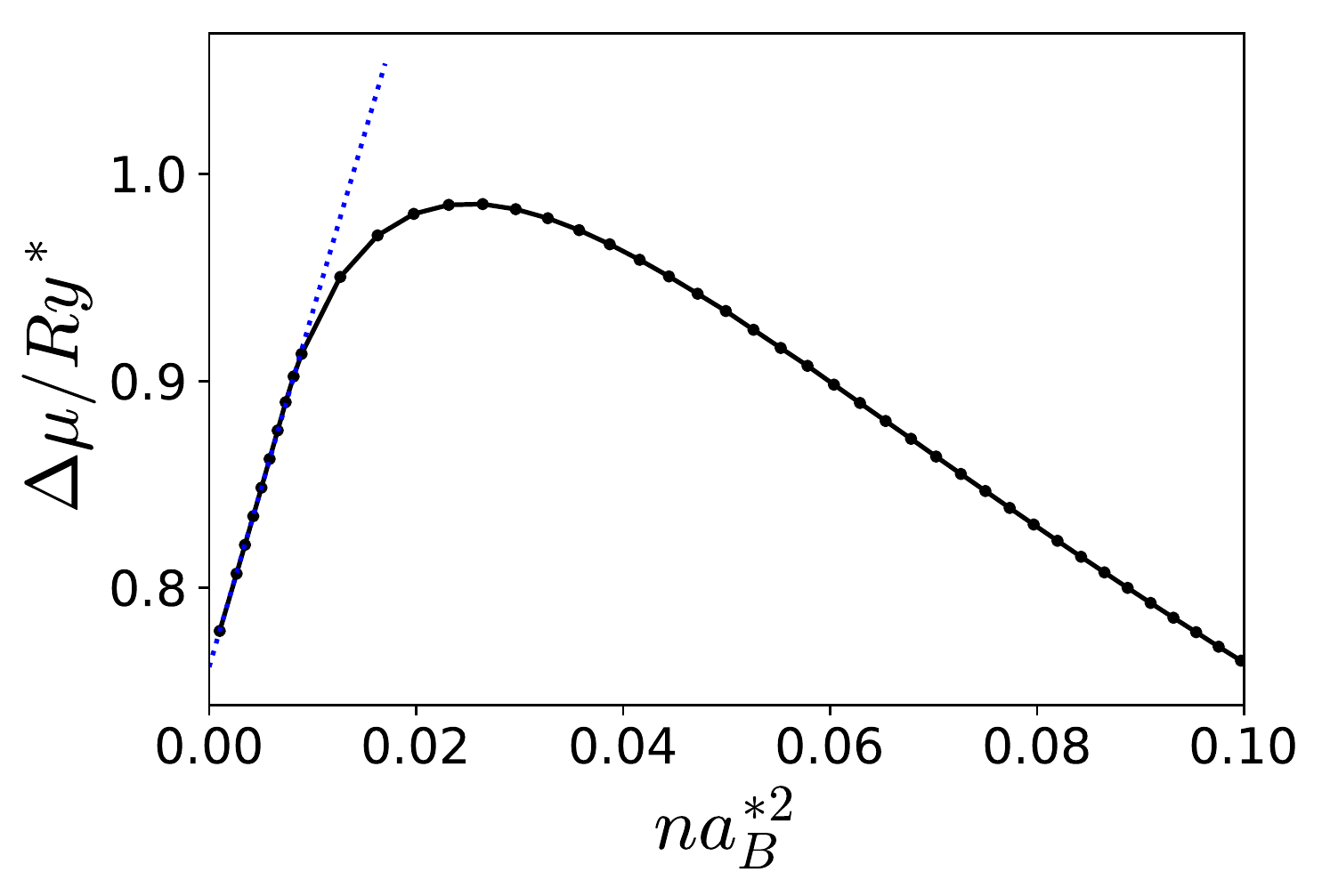}
\subcaption{} \label{fig:dmu_nx}
\end{subfigure}
\caption{Chemical potentials as functions of carrier densities from mean-field calculations: (a) exciton chemical potential $\mu_x$ vs. carrier density $n(=p)$; (b) chemical potential of electrons $\mu_e$ vs. electron density $n$ at fixed $p$; (c) chemical potential jump $\Delta\mu$ at neutrality vs. carrier density $n(=p)$.}
\end{figure*}

We consider a bilayer electron-hole system with electron and hole densities 
$n$ and $p$, neutralizing backgrounds in both electron and hole layers,
and the band extrema energies of both the conduction band and the valence band set to zero.
With these conventions, the electron and hole chemical potentials 
of the interacting electron-hole system correspond to the chemical potentials 
$\mu_e(n,p)$ and $\mu_h(n,p)$ in Eqs.~\eqref{eq:equil}.
The mean-field Hartree-Fock Hamiltonian of the electron-hole system is
\begin{equation}
H_{\textrm{MF}} = H_{eh} + H_X
\end{equation}
where $H_{eh}$ is the two-band single-particle Hamiltonian for electrons and holes
\begin{equation}
H_{eh} = \sum_{\bm k} (\frac{\hbar^2k^2}{2m_e^*}a_{c\bm k}^{\dagger}a_{c\bm k}-\frac{\hbar^2k^2}{2m_h^*}a_{v\bm k}^{\dagger}a_{v\bm k})
\end{equation}
where $c/v$ stand for the conduction band of the electron layer and the 
valence band of the hole layer. 
As explained above, the zero of energy for the electron layer is chosen at the bottom of the conduction 
band, and for the hole layer at the top of the valence band. 
Because of the assumption of neutralizing backgrounds there is no electrostatic mean-field potential,
and $H_X$ is the exchange self-energy:
\begin{equation}
\label{eq:hf}
H_X = -\frac{1}{A}\sum_{\bm k \bm k'} V_{b'b}(\bm k'-\bm k)\rho_{b'b}(\bm k')a_{b'\bm k}^{\dagger}a_{b\bm k}
\end{equation}
where $A$ is the area of the two-dimensional system, $b=c,v$ is the band index,
$V_{cc}(q)=V_{vv}(q)=V(q)=2\pi e^2/\epsilon q$ is the intralayer Coulomb interaction,
$V_{cv}(q)=V_{vc}(q)=U(q)=V(q)\exp(-qd)$ is the interlayer Coulomb interaction at interlayer distance $d$,
and $\epsilon$ is the dielectric constant of the surrounding three-dimensional material.
In Eq.~\eqref{eq:hf}
$\rho$ is the density matrix calculated relative to the density matrix 
when the conduction band is empty and the valence band is full:
\begin{equation}
\rho_{b'b}(\bm k) = \langle a_{b\bm k}^{\dagger}a_{b'\bm k}\rangle-\delta_{b'b}\delta_{bv}
\end{equation}

Below we express lengths and energies in terms of characteristic scales $a_B^* = \epsilon\hbar^2/m^*e^2$ and $Ry^* = e^2/2\epsilon a_B^*$, where $m^*=m_e^*m_h^*/(m_e^*+m_h^*)$ is the reduced effective mass. For simplicity we perform numerical calculations only for the $m_e^*=m_h^*=2m^*$ case. 
This model approximates MoSe$_2$/WSe$_2$ bilayers surrounded by hexagonal boron nitride (hBN) dielectrics if we 
choose $m^*\approx 0.2\,m_e$, and $\epsilon\approx 5$, which implies that 
$a_B^*\approx\SI{1.3}{nm}$ and $Ry^*\approx\SI{0.11}{eV}$. 
The calculations below are for the case $d=a_B^*$, corresponding approximately to the case 
with one or two hBN layers between the MoSe$_2$ and WSe$_2$ monolayers.

In order to provide an interpretation for
the mean-field theory results presented below, we first discuss  
a simple phenomenological picture in which we assume that the electron-hole fluid 
consists of paired (excitons) and unpaired electrons or holes.
If we assume that the physically preferred state of the system is one with the maximum number of 
electron-hole pairs, then the system consists of $n_x=\min(n,p)$ excitons and $|n-p|$ unpaired electrons or holes. 
In this picture the total energy per area of the system at small $n$ and $p$ can be 
approximated by the simple expression
\begin{equation} \label{eq:E_np}
\varepsilon(n,p) = - \varepsilon_{b} n_x + \frac{g}{2} n_x^2 + \frac{\alpha}{2}(n-p)^2
\end{equation} 
where $\varepsilon_b$ is the exciton binding energy, $g$ describes exciton-exciton interactions, and $\alpha$ is the inverse thermodynamic density-of-states of the unpaired electrons or holes. This expression implies that the total energy 
$\varepsilon(n,p)$ has a cusp at $n=p$.  The chemical potentials in which we are primarily interested are given by  
\begin{equation} 
\begin{split} \label{eq:mu_def}
\mu_e(n,p) &= \frac{\partial \varepsilon(n,p)}{\partial n} \\
\mu_h(n,p) &= \frac{\partial \varepsilon(n,p)}{\partial p}.
\end{split}
\end{equation}
For $n<p$ adding an electron adds an exciton and removes a free carrier whereas for $n>p$ adding an electron simply adds a free 
carrier.  In a similar way adding a hole simply adds a free carrier for $n<p$, whereas for $n>p$ it adds an exciton and removes a free carrier.  For this reason the chemical potentials of electrons and holes jump in opposite directions 
by $\varepsilon_{b}$ when the $n=p$ line is crossed.  The chemical potential for excitons
\begin{equation} \label{eq:mux_def}
\mu_x = \mu_e+\mu_h
\end{equation}
is however continuous.  For $n=p=n_x$ our {\it ansatz} energy expression, Eq.~\eqref{eq:E_np},  
places all electrons and holes into pairs and the exciton chemical potential 
\begin{equation} \label{eq:mux_nx}
\mu_x(n_x) = -\varepsilon_b + g n_x.  
\end{equation}

We now compare this simple physical picture with microscopic mean-field theory calculations
performed as a function of $n$ and $p$.
First of all, in Fig.~\ref{fig:mux_n=p} we plot the exciton chemical potential at 
$n=p$ calculated as a function of $n_x$ and compare it with Eq.~\eqref{eq:mux_nx}. 
By linearly fitting the small-$n$ part of the curve we find that the exciton binding 
energy $\varepsilon_b=0.77\,Ry^*$ and the exciton-exciton interaction parameter 
$g=-12\,(Ry^* a_B^{*2})$. The negative value of $g$ here implies that the interaction between
interlayer excitions would be attractive at the considered interlayer separation $d = a_B^*$
if electrostatic interactions, which are suppressed at this stage because 
of the neutralizing backgrounds in each layer in the reference system calculation,
were ignored.  Technically this property reflects the dominance of intralayer Coulomb interactions,
and agrees with previous studies \cite{wu2015theory}. 
At large exciton density ($n_xa_B^{*2} \gtrsim 0.02$) higher-order 
interaction terms come into play, the curve deviates from the linear relation of Eq.~\ref{eq:mux_nx},
and the simple expression \eqref{eq:E_np} is no longer a good approximation.

According to Eq.~\eqref{eq:E_np}
\begin{equation}
\mu_e = \alpha (n-p) - (\varepsilon_b-gn)\Theta(p-n)
\end{equation}
where $\Theta(x)$ is the Heaviside step function. 
The electron chemical potential calculated from microscopic mean-field theory is plotted as a function of $n$ for fixed $p$ in Fig.~\ref{fig:mue_n}. 
The upward jump in the electron chemical potential 
anticipated in the ansatz expressions is prominent.  The parameter $\alpha$, the quasiparticle inverse thermodynamic density of states, accounts for the rate of change of electron chemical potential near the $n=p$ point.
A linear fit gives $\alpha=-115\,(Ry^* a_B^{*2})$.  This negative value of $\alpha$ is 
related to the negative compressibility property, which applies also in single-component systems, and is due
to the dominance of exchange interactions in low-density electron gases \cite{eisenstein1992negative,eisenstein1994compressibility,ying1995quantitative}.  
The size of the chemical potential jump at $n=p$ is plotted vs. carrier density in Fig.~\ref{fig:dmu_nx}.
The ansatz expression for this quantity is
\begin{equation}
\Delta\mu(n_x) = \varepsilon_b - gn_x = -\mu_x.
\end{equation}
However, comparison of Fig.~\ref{fig:mux_n=p} and \ref{fig:dmu_nx} shows that the behavior of $-\mu_x$ and $\Delta\mu$ are very 
different when $n$ is not small.  
Fitting these two quantities at the small-$n$ part yields different slopes but almost identical $n \to 0$ intercepts.
This implies that while the physical picture represented by Eq.~\eqref{eq:E_np} is oversimplified, it does capture
a large part of the truth.

The chemical potentials calculated in this section are based on a temperature $T=0$ mean-field theory, which 
yields an electron-hole pair condensate ground state.  Although this theory is expected to become exact at $T=0$ in the limit of 
very low carrier densities, it is expected to fail at large carrier densities.  It is generally thought 
that \cite{fogler2014high,perali2013high,neilson2014excitonic} 
at high enough carrier density a Mott limit is reached above which exciton condensation does not 
occur.  Although we use these chemical potentials in the following section to discuss some qualitative expectations,
an equally important aspect of our analysis is that it can be used in combination with experimental measurements of the 
dependence of carrier densities on gate voltages to extract chemical potential data from experimental observations.
Experiments of this type would be especially valuable at higher temperatures where mean-field theory is not expected to be 
reliable.  

\section{Gate Control of Carrier Densities} \label{sec:control}

\begin{figure}
\centering
\includegraphics[width=\linewidth]{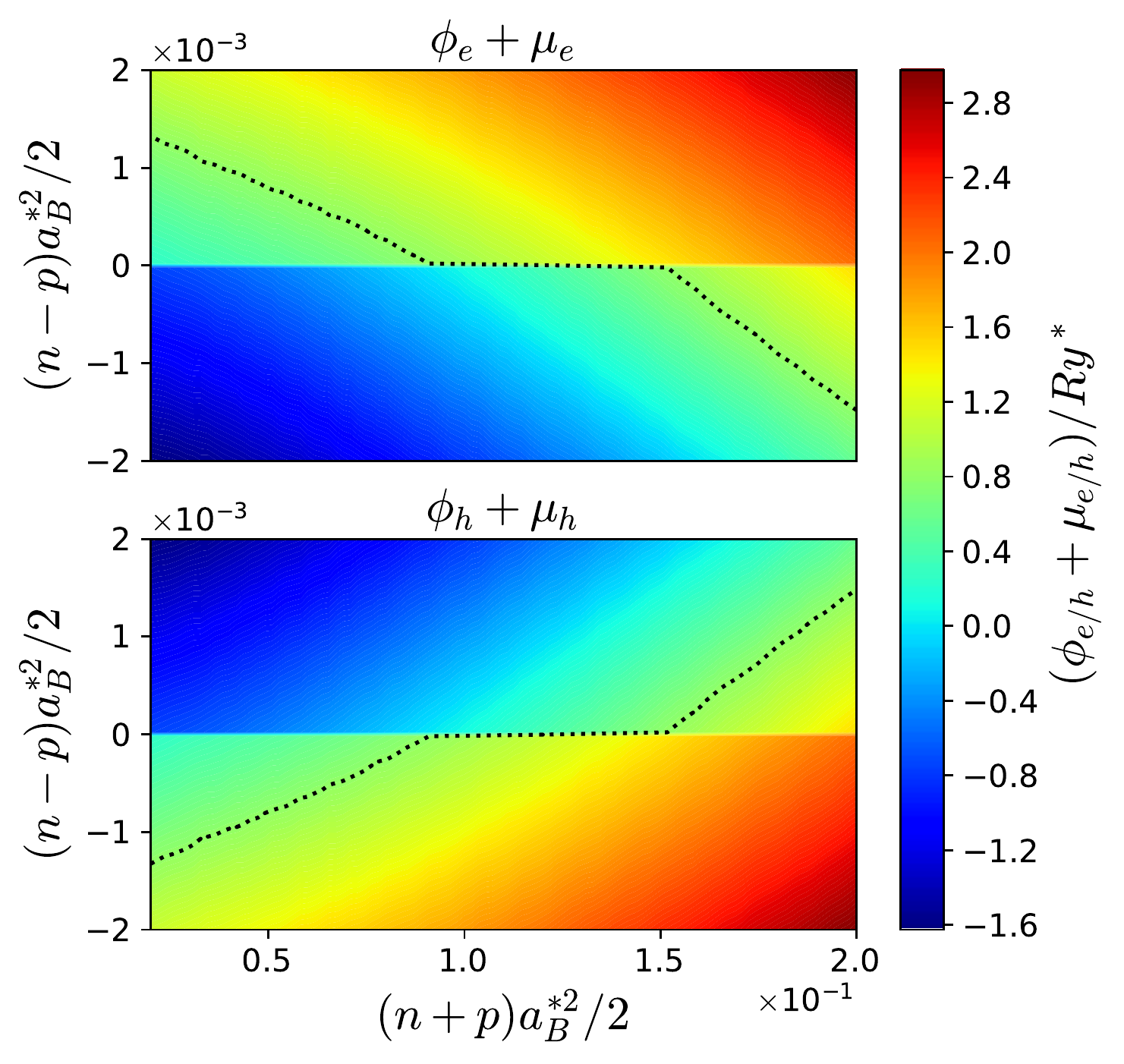}
\caption{Color plot of $\phi_e+\mu_e$ (upper panel) and $\phi_h+\mu_h$ (lower panel) as functions of $(n+p)/2$ and $(n-p)/2$. The black dotted lines are contour lines along which $\phi_{e/h}+\mu_{e/h}=0.8Ry^*$. For the calculation of $\phi_e$ and $\phi_h$ we assume $d_e=d_h=20a_B^*$ and $d_s=a_B^*$.} \label{fig:V_np}
\end{figure}

The left-hand sides of Eqs.~\eqref{eq:equil} are functions of $n$ and $p$.  
For given values of $n$ and $p$, $\phi_e$ and $\phi_h$ are given by the electrostatic equations \eqref{eq:phi_np}, 
while $\mu_e$ and $\mu_h$ are more complex quantities that depend on correlations in the 
electron-hole fluid.  
For given gate voltages $V_e$ and $V_h$, Eqs.~\eqref{eq:equil} can have one or more solutions, the latter 
instance would signal bistability or multi-stability of the electron-hole system.
In Figs.~\ref{fig:V_np} we plot $\phi_{e/h}+\mu_{e/h}$, 
calculated using mean-field theory chemical potentials,
as functions of total carrier density $(n+p)/2$ and total charge density $(n-p)/2$. 
One important feature of in this plot is the jump in electrochemical potentials across the $n=p$ line discussed above.
Aside from the jump at neutrality, increasing net charge density changes the electrochemical 
potentials rapidly mainly because of the electrostatic contributions.  $\phi_{e}$ increases with $n-p$, 
$\phi_{h}$ decreases with $n-p$, and both changes are rapid because of the relatively large distance to the gates.
(Note the difference in scales for the $x$ and $y$ axes in Figs.~\ref{fig:V_np}.)
At fixed $n-p$, both electrochemical potentials increase with $n+p$, mainly because of band-filling.

From Eqs.~\eqref{eq:equil} we see that the equilibrium values of $n$ and $p$ for given values of the 
control gate voltages $V_e$ and $V_h$ correspond to the intersections of the contour 
lines in Figs.~\ref{fig:V_np} along which 
$\phi_e(n,p) + \mu_e(n,p)=V_e-E_c$ and $\phi_h(n,p) + \mu_h(n,p)= V_h-E_v$.
Since the contour lines of $\phi_e + \mu_e$ have negative slopes when plotted with 
$n+p$ along the $x$-axis and $n-p$ along the $y$-axis (as in Figs.~\ref{fig:V_np}) while those for 
$\phi_h + \mu_h$ have positive slopes, the lines never intersect more than once. 
We conclude the bistability is unlikely when the electron and hole layers are separated by an 
hBN tunneling barrier and the gates are well separated from the active layers.

\begin{figure}
\centering
\includegraphics[width=0.8\linewidth]{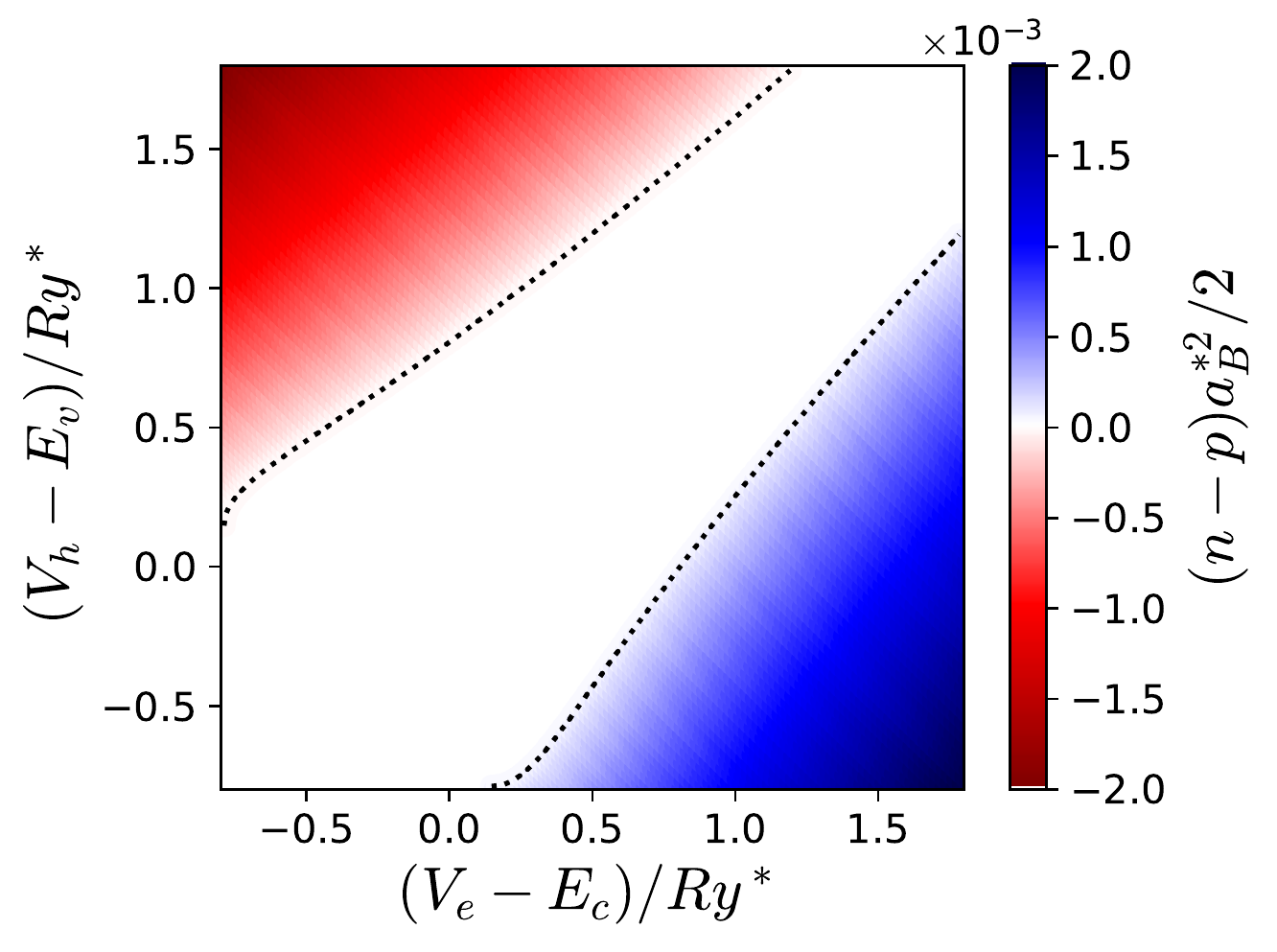}
\caption{Color plot of $(n-p)/2$ as a function of $V_e-E_c$ and $V_h-E_v$. The black dotted lines are the boundaries of the $n=p$ region.} \label{fig:n-p_V}
\end{figure}

\begin{figure}
\centering
\includegraphics[width=\linewidth]{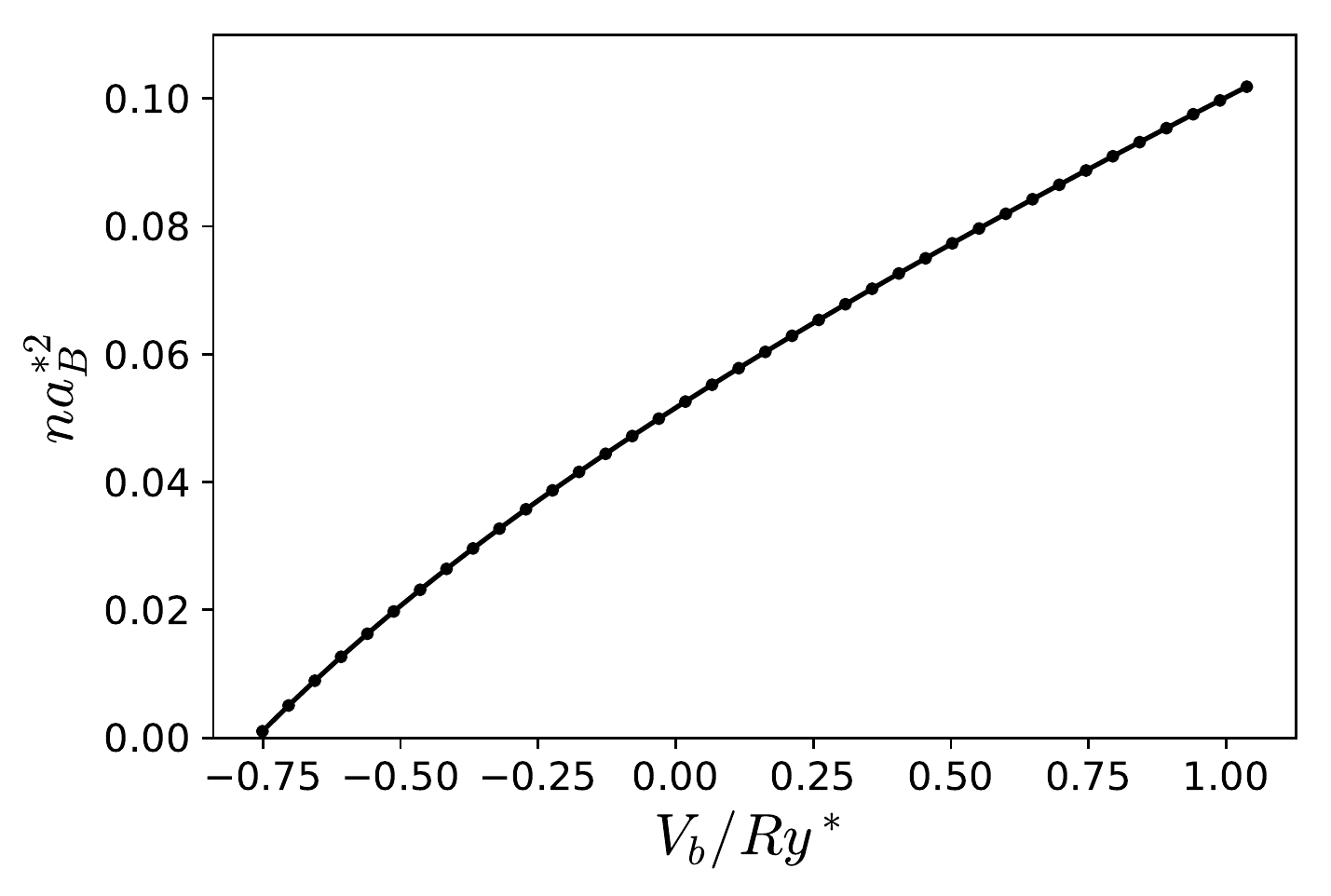}
\caption{Exciton density $n=p=n_x$ as a function of bias voltage $V_b$ 
when $(V_e,V_h)$ is in the neutral region.} \label{fig:n_Vb}
\end{figure}

\begin{figure}
\centering
\includegraphics[width=0.8\linewidth]{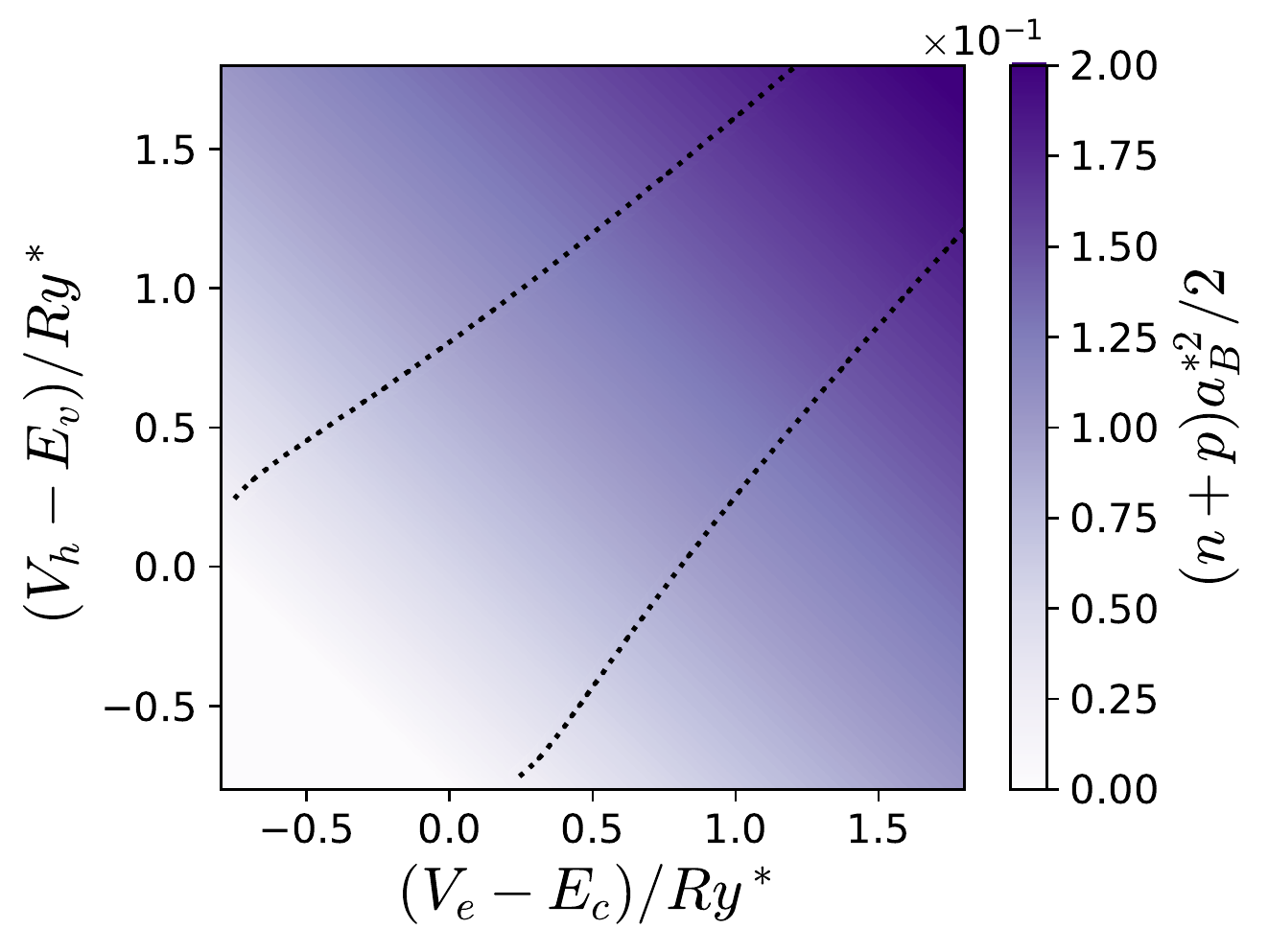}
\caption{Color plot of $(n+p)/2$ as a function of $V_e-E_c$ and $V_h-E_v$. The black dotted lines are the boundaries of the $n=p$ region.
The total carrier density is a smooth function of the voltage control parameters.} \label{fig:n+p_V}
\end{figure}

Special care must be taken along the $n=p$ line where the electron and hole 
electrochemical potentials jump, yielding 
the horizontal portions of the contour lines in Fig.~\ref{fig:V_np}.
The electron and hole chemical potentials are ill defined along these horizontal line segments 
because it is not energetically allowed to exchange particles with individual reservoirs.  
It is however possible to exchange an electron-hole pair with the reservoirs
via a two-particle correlated 
tunneling process \cite{xie2018electrical}.
Correlated electron-hole tunneling is relevant when the difference between $V_e-E_c$ and $V_h-E_v$ 
is smaller than the chemical potential jump at neutrality, so that the horizontal segments of two contour lines 
in Fig.~\ref{fig:V_np} intersect.
Under this circumstance the exciton chemical potential is still 
well defined, and its value at $n=p=n_x$ is determined by solving 
the pair equilibrium condition
\begin{equation}
\label{eq:ex_equil}
\phi_{e}+\mu_{e}+ \phi_{h}+\mu_{h} = V_e -E_c + V_h - E_v.  
\end{equation}
This implies a finite region (rather than a line) in the $(V_e,V_h)$ parameter space in 
which $n=p$ as illustrated in Fig.~\ref{fig:n-p_V}. The width of this region first increases with increasing gate voltages and 
then narrows.  Defining the right-hand side of Eq.~\eqref{eq:ex_equil} as the bias voltage $V_b$  
we obtain the results for the exciton density as a function of $V_b$ shown in Fig.~\ref{fig:n_Vb}. 
The system is supplied with excitons when $V_b > - \varepsilon_b$.  Above the bias voltage threshold  the curve is initially linear because of exciton-exciton interactions.  The sum of electron and 
hole densities varies smoothly with gate voltage in the entire $(V_e,V_h)$ plane as illustrated 
in Fig.~\ref{fig:n+p_V}.

\section{Leakage Currents} \label{sec:leak}

The theory in the previous sections applies when there is no current flowing between layers.
In the presence of a tunneling current, the theory still applies as a good approximation if the current is small enough
so that we can assume quasi-equilibrium in each layer. In the following we discuss the opposite 
limit, in which a large leakage current between layers requires a non-equilibrium transport picture. 
By analyzing this limit we are able to qualitatively explain the transport characteristics observed
in a recent experimental study of electrically controlled electron-hole bilayers \cite{wang2019evidence}.

\begin{figure}
\centering
\includegraphics[width=\linewidth]{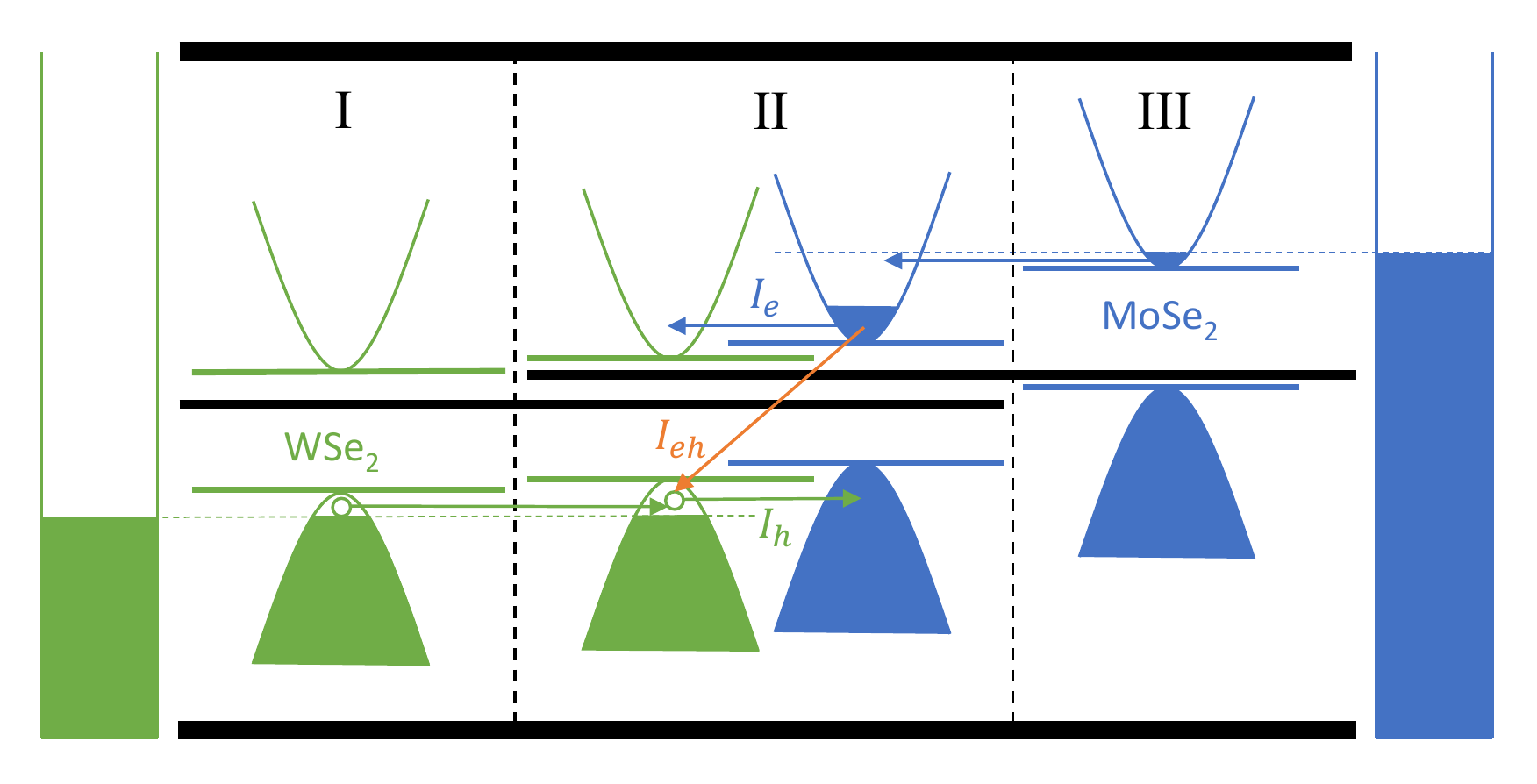}
\caption{Schematic of electron-hole transport processes in the bilayer system with isolated monolayer regions near the contacts
and effective contact resistances that are larger than effective interlayer resistances.}
\label{fig:transport}
\end{figure}

For definiteness we consider MoSe$_2$/WSe$_2$ bilayer systems in which 
the MoSe$_2$ layer is n-type and the WSe$_2$ is p-type. In a typical experimental setup, the two layers 
do not perfectly overlap, leaving isolated monolayer regions between the bilayer region and the 
electrical contacts. We sketch an idealized geometry in Fig.~\ref{fig:transport} 
in which we divide the system into three parts: regions I and III are isolated WSe$_2$ and MoSe$_2$ monolayer regions, 
while region II is the overlapped bilayer region. Assuming that 
$V_h$ is above threshold, region I is filled with holes with density $p_{\rm{I}}\propto V_h-E_v$. 
Imagine we now increase $V_e$ from a small value. When $V_e$ is below threshold, there are no free carriers in the MoSe$_2$ layer, so no current flows and the electric field distribution is not affected by the MoSe$_2$ layer. The whole WSe$_2$ layer is uniformly charged and in equilibrium with the left lead. The electric potential on the MoSe$_2$ layer in region II is very close to that of the WSe$_2$ layer due to the small interlayer distance, while in region III the electric potential is the same as the gates since the whole layer is neutral. 
The electric potential jump between regions II and III implies a sharp bending of the MoSe$_2$ bands at the boundary.

When $V_e$ increases above threshold, region III starts to get charged toward 
electron density $n_{\rm{III}}\propto V_e-E_c$. Due to the electric potential jump, current is injected into empty states in region II and tries to fill the band to the same level as in region III. The injection current $I_{\rm{e,inj}}$ is a function of $n_{\rm{III}}$ (and therefore depends on $V_e-E_c$ only).  Importantly it does not depend on the physics in regions I and II (as long as region II is far from equilibrium with III). Were it not for the leakage current between the two layers in region II, the whole MoSe$_2$ layer would finally come to equilibrium with the right lead, and the theory in the previous sections would apply. 
However, as we schematically show in Fig.~\ref{fig:transport}, a large leakage current can prevent the system from 
establishing equilibrium. The leakage current $I_{\rm{leak}}$ is a monotonically increasing function of the carrier densities 
$n_{\rm{II}}$ and $p_{\rm{II}}$.  If $n_{\rm{II}}$ and $p_{\rm{II}}$ reach value such that $I_{\rm{leak}}=I_{\rm{e,inj}}$ before equilibrium is established, the system will reach a non-equilibrium current-carrying steady state. 
The current through the system is limited by the injection current from region III, which depends only on $V_e-E_c$. 
Meanwhile, the WSe$_2$ layer in regions I and II maintains equilibrium with the left lead.

Similar considerations apply to region I, which like region III also has a maximum injection current $I_{\rm{h,inj}}$ that is a function of $p_{\rm{I}}\propto V_h-E_v$. 
When $V_e$ is so large that $I_{\rm{e,inj}}>I_{\rm{h,inj}}$, the tunneling current is limited by the injection current from region I, while regions II and III stay in equilibrium with the right lead. The transition takes place when $I_{\rm{e,inj}}(n_{\rm{III}}) = I_{\rm{h,inj}}(p_{\rm{I}})$. Since $I_{\rm{e,inj}}(n_{\rm{III}})$ and $I_{\rm{h,inj}}(p_{\rm{I}})$ are similar functions of carrier density, roughly 
speaking this happens when $n_{\rm{III}}=p_{\rm{I}}$, or $V_e-E_c=V_h-E_v$.
The most important conclusion from this analysis above is that, as long as the system is in the p-n regime, the tunneling current is determined by the following equation:
\begin{equation}
I(V_e,V_h) = \min\{I_{\rm{e,inj}}(V_e), I_{\rm{h,inj}}(V_h)\}
\end{equation}

The leakage current consists of three parts (see Fig.~\ref{fig:transport}): electron tunneling $I_e$, hole tunneling $I_h$, and interband electron-hole phonon-assisted tunneling or optical recombination $I_{eh}$. 
Due to the relatively weak electroluminescence observed in experiments, we assume that $I_{eh}\ll I_e,I_h$.
Then the tunneling current has a threshold when the chemical potential in the electron layer is above the bottom of the conduction band in the hole layer (or when the chemical potential in the hole layer is below the top of the valence band in the electron layer). Above threshold the tunneling conductance far exceeds the corresponding injection current conductance from monolayer regions at the same chemical potential. 
In this case the total current is limited by injection from monolayer regions as discussed above, and the details of the interlayer tunneling process do not matter. One possible way to study interlayer tunneling is to tune the chemical potential below threshold, so that the tunneling current only has a single weak component $I_{eh}$. This can be made easier by applying a vertical electric field between the two layers to tune the indirect band gap. The dependence of $I_{eh}$ on carrier densities $n,p$ can provide clues that help us 
understand the correlated electron-hole states and possibly provide evidence for exciton condensation.

We now provide a crude estimate of the injection current by considering the ballistic transport limit. 
Consider an electron with momentum $(k\cos\theta, k\sin\theta)$ moving across a rectangular region with length $L$ and width $W$. 
The electron has velocity $\hbar k\cos\theta/2m_e^*$ along $L$, and contributes to current
\begin{equation}
i(k,\theta) = e/\frac{L}{\hbar k\cos\theta/m_e^*} = \frac{e\hbar k\cos\theta}{m_e^* L}
\end{equation}
Assuming that all electrons with $k_x>0$ contribute to the current, we integrate over half of the Fermi
sea (accounting for valley degeneracy) to obtain 
\begin{equation} \label{current}
I = 2\int_{-\pi/2}^{\pi/2}d\theta\int_{0}^{k_F}\frac{kdk}{(2\pi)^2/LW}i(k,\theta) = \frac{e\hbar Wk_F^3}{3\pi^2m_e^*}.
\end{equation}
Here the Fermi momentum is related to the carrier density by
\begin{equation}
\frac{\pi}{2}k_F^2 = nWL\frac{(2\pi)^2}{2WL} = 2\pi^2 n \;\Rightarrow\; k_F^2 = 4\pi n
\end{equation}
For the monolayer region, it is a good approximation to assume $\phi=V_e-E_c$ when $V_e>E_c$ because of its large distance from the gates. 
By simple electrostatics the carrier density is given by
\begin{equation}
\frac{4\pi ne^2}{\epsilon} = \phi\;(\frac{1}{d_e}+\frac{1}{d_h})
\end{equation}
For $\phi=\SI{1}{eV}$, $d_e=d_h=\SI{25}{nm}$ and $\epsilon=5$, we obtain $n=\SI{2.21e12}{cm^{-2}}$. 
Then for $W=\SI{3}{\micro\meter}$ and $m_e^*=0.4\,m_e$, from Eq.~\eqref{current} we estimate the current
\begin{equation}
I = \SI{687}{\micro\ampere},
\end{equation}
which is, as expected, much larger than the current observed in experiment due to the unrealistic assumption of perfect transmission. 
The ballistic transport picture gives $I\propto(V_e-E_c)^{3/2}$, in contrast to the $I\propto(V_e-E_c)^3$ relation found in experiment. The difference can be explained by the fact that electron transmission is not perfect and depends on energy. We can make our result agree with the experiment by introducing a phenomenological energy-dependent transmission probability $T(E)\propto E^{3/2}$ where $E$ is the electron energy relative to conduction band bottom. With a similar analysis for holes, we restore the cubic law found in experiment \cite{wang2019evidence}:
\begin{equation}
I(V_e,V_h) \propto \min\{V_e-E_c,V_h-E_v\}^3. 
\end{equation}

\section{Discussion} \label{sec:discussion}

In this article we have analyzed how two gate voltages $V_e$ and $V_h$ can be used to control 
the electron and hole densities $n$ and $p$, 
and hence all properties, of bilayer two-dimensional electron-hole fluids.  
When there is no leakage current between electron and hole
layers contacted by electrodes with different chemical potentials, bias voltages can be used to tune the bilayer 
into an effective semi-metal regime in which carriers are present in both $n$ and $p$ layers.  Because of jumps in electron and hole 
chemical potentials related to electron-hole pair bound state formation that occur along the $n=p$ carrier-compensation line,
there is a finite area in the $(V_e,V_h)$ gate-control region, instead of a line, within which $n=p$.  
In this region the electron and hole chemical potentials are not individually well defined.  
Instead their sum, identified as the exciton chemical potential, is well defined and controlled by the value of $V_e+V_h$.  Fig.~\ref{fig:n-p_V} outlines the boundaries 
of the excitonic area in the control space as estimated by mean-field theory.  
We anticipate that the width of this region, which is a measure of the chemical potential 
jumps at neutrality - referred to as the exciton binding energy below - 
narrows with increasing exciton density due to screening \cite{fogler2014high,perali2013high,neilson2014excitonic}, exciton-exciton attractive van der Waals interactions \cite{lozovik1997phase}, 
and other effects that are not captured by mean-field theory.  Although the fate of this 
quantity, which has not previously been experimentally accessible, is usually not explicitly 
discussed in the literature, the most common view 
appears to be that the exciton binding energy drops suddenly to zero at 
a first-order Mott transition.  Another view is that the exciton binding energy declines 
monotonically with increasing density in a crossover between BEC and BCS limits, but never vanishes.
Still another possiblity is that the exciton binding energy vanishes smoothly at a continuous 
phase transition.  Capacitive studies of electron-hole bilayers should be able to provide a 
definitive experimental answer to this question.

Within the finite-area exciton region of the gate control space $V_e+V_h$ acts as a 
chemical potential for excitons.  By using contact pairs to establish exciton reservoirs with different chemical 
potentials at different locations in a bilayer system, we anticipate that it will be possible to 
measure electrical and thermal transport properties of excitons, and to design exciton circuits.
This exciting prospect is a principle motivation for formulating the analysis in this Article.  

We have compared our analysis with a recent paper that reports on important progress toward 
electrical control of electron-hole bilayers, and therefore of the their optical properties.
We conclude that the transport properties reported in Ref.~\onlinecite{wang2019evidence} 
are indicative of electron-hole systems that are not in a quasi-equilibrium configuration, 
but instead in a transport steady state that is strongly influenced by 
the intra-band leakage current between the bilayers.  The leakage current is easily reduced, however, by inserting 
dielectric layers between the 2D semiconductors, or simply by reducing the area of overlap between 
the two 2D semiconductor layers.  We can anticipate that further progress will enable 
accurate control of quasi-equilibrium electron-hole bilayers, and by staying within the $n=p$ region, 
electrical and thermal control of exciton currents.

The relationship between electron and hole densities and gate voltages depends on both many-body 
correlations and disorder 
within the 2D semiconductor layers.  
In this paper we have ignored disorder and used a mean-field theory to
describe interactions in the electron-hole system.  
The mean-field theory is expected to be accurate at low electron and hole densities, 
but it is precisely in this regime that disorder is most important.  
We expect that future experimental studies of electrically 
controlled electron-hole bilayers, and future progress controlling disorder, will shed light on both 
aspects of electron-hole bilayer physics. 

\section{Acknowledgements} 
The authors thank Kin Fai Mak and Jie Shan for helpful discussions. This work was supported by the U.S. Department of Energy, Office of Science, Basic Energy Sciences, under Award \# DE‐SC0019481
and by Welch Foundation grant TBF1473.  

\bibliography{ehfluid.bib}

\end{document}